\documentclass[aps,prl,twocolumn,showpacs,superscriptaddress,nofootinbib,preprintnumbers]{revtex4-2}%

\usepackage{graphicx}
\usepackage{tabularx}
\usepackage{bm}
\usepackage{latexsym}
\usepackage{epsf}
\usepackage{rotating}
\usepackage{epsfig,graphics,rotate,color}
\usepackage{wrapfig}
\usepackage{amssymb}
\usepackage{amsmath}
\usepackage{amsfonts}
\usepackage{array,hhline,dcolumn}%
\usepackage[normalem]{ulem}
\usepackage{color}
\usepackage{style}
\usepackage{siunitx}
\usepackage{placeins}
\newcommand{\mycomment}[1]{}

\DeclareSIUnit\year{yr}

\begin{document}
\bibliographystyle{apsrev4-1}

\title{The Final Frontier for Proton Decay}

\author{Sebastian Baum \,\orcidlink{0000-0001-6792-9381}}
\email{sbaum@physik.rwth-aachen.de}
\affiliation{Institute for Theoretical Particle Physics and Cosmology, RWTH Aachen University, D-52056 Aachen, Germany}

\author{Cassandra Little \,\orcidlink{0000-0000-0000-0000}}
\email{littleca@umich.edu }
\affiliation{University of Michigan, Ann Arbor, Michigan 48109, USA}

\author{Paola Sala \,\orcidlink{0000-0001-9859-5564}}
\email{paola.sala@orange.fr}
\altaffiliation[]{Retired}
\affiliation{INFN Milano, via Celoria 16, I-20133 Milano, Italy}

\author{Joshua Spitz \,\orcidlink{0000-0002-6288-7028}}
\email{spitzj@umich.edu}
\affiliation{University of Michigan, Ann Arbor, Michigan 48109, USA}

\author{Patrick Stengel \,\orcidlink{0000-0002-1000-0050}}
\email{pstengel@fe.infn.it}
\affiliation{INFN Ferrara, via Giuseppe Saragat 1, I-44122 Ferrara, Italy}

\preprint{TTK-24-22}

%--------------------------------------------------------------------
\begin{abstract}
We present a novel experimental concept to search for proton decay. Using paleo-detectors, ancient minerals acquired from deep underground which can hold traces of charged particles, it may be possible to conduct a search for $p \to \bar{\nu} K^+$ via the track produced at the endpoint of the kaon. Such a search is not possible on Earth due to large atmospheric-neutrino-induced backgrounds. However, the Moon offers a reprieve from this background, since the conventional component of the cosmic-ray-induced neutrino flux at the Moon is significantly suppressed due to the Moon's lack of atmosphere. For a 100\,g, $10^9$\,year old (100\,kton$\cdot$year exposure) sample of olivine extracted from the Moon, we expect about 0.5 kaon endpoints due to neutrino backgrounds, including secondary interactions. If such a lunar paleo-detector sample can be acquired and efficiently analyzed, proton decay sensitivity exceeding $\tau_p\sim10^{34}$\,years may be achieved, competitive with Super-Kamiokande's current published limit ($\tau_p>5.9\times 10^{33}$\,years at 90\% CL) and the projected reach of DUNE and Hyper-Kamiokande in the $p \to \bar{\nu} K^+$ channel. This concept is clearly futuristic, not least since it relies on extracting mineral samples from a few kilometers below the surface of the Moon and then efficiently scanning them for kaon endpoint induced crystal defects with sub-micron-scale resolution. However, the search for proton decay is in urgent need of a paradigm shift, and paleo-detectors could provide a promising alternative to conventional experiments.
\end{abstract}
\maketitle
%--------------------------------------------------------------------

%--------------------------------------------------------------------
\section{Introduction}
%--------------------------------------------------------------------

The proton is the lightest baryon in the Standard Model (SM) of particle physics. The SM's gauge structure and particle content render baryon number ($B$) an accidental symmetry of the SM. While non-perturbative processes in the SM do violate baryon number~\cite{tHooft:1976rip}, these {\it electroweak instantons} are characterized by $\Delta B = 3$ rather than $\Delta B = 1$ as in proton decay, rendering protons absolutely stable in the SM. Since the stability of the proton in the SM is accidental, it is natural to ask whether protons decay in models of physics Beyond the SM (BSM). Indeed, proton decay is a hallmark of many well-motivated classes of BSM models, especially in Grand Unified Theories (GUTs). While the precise proton lifetime depends on the particle content and symmetry-breaking scheme, SU(5)-based GUT models tend to predict proton lifetimes $\tau_p \lesssim 10^{34}\,$yr~\cite{Georgi:1974sy,Dimopoulos:1981zb,Sakai:1981pk,Nath:1985ub} (although longer lifetimes are possible~\cite{Hebecker:2002rc,Ellis:2002vk,Pati:2003qia,Arkani-Hamed:2004zhs,Alciati:2005ur,Dorsner:2005fq}) and models based on SO(10) tend to predict somewhat larger lifetimes, $\tau_p \sim 10^{32} - 10^{40}\,$yr~\cite{Fritzsch:1974nn,Lucas:1996bc,Babu:1997js,Shafi:1999vm,Pati:2003qia,Dorsner:2005fq,King:2021gmj}. In general, protons could decay in a number of final states. Broadly speaking, $p \to e^+ \pi^0$ tends to be the most important decay channel in non-supersymmetric GUT models, while $p \to \bar{\nu} K^+$ is the leading decay mode in many supersymmetric GUT models.

The search for proton decay has been a central pursuit of the particle physics community over the last 50 years because it probes physics at extremely high energy scales and, in particular, might be the only signature of GUT models accessible by conceivable laboratory experiments. The first dedicated experiments were built in the 1950s and '60s, using few-hundred kg scale detectors and establishing limits on the proton lifetime of $\tau_p > 10^{29}\,$years~\cite{Perkins:1984rg}. Subsequent generations of proton-decay experiments used increasingly large detectors from the ton to the kton scale~\cite{Cherry:1981uq,Bartelt:1982vr,Battistoni:1983ka,Irvine-Michigan-Brookhaven:1983iap,Arisaka:1985lki,Kajita:1986gk,Bartelt:1986cv,FREJUS:1987krx,FREJUS:1988pvk,Kamiokande-II:1989avz,Thron:1989cd,Soudan-2:1998blg,Becker-Szendy:1992ufh,McGrew:1999nd}. In the 1990s Super-Kamiokande, a water Cherenkov detector with ${\sim\,}25\,$kton fiducial volume, was constructed. After more than 20 years of operations, Super-Kamiokande is currently providing the most stringent constraints on the proton lifetime: $\tau(p \to e^+ \pi^0) > 2.4 \times 10^{34}\,$yr at 90\,\% confidence limit (CL) using 450\,kton$\cdot$year of data~\cite{Super-Kamiokande:2016exg,Super-Kamiokande:2020wjk}, and $\tau(p \to \bar{\nu} K^+) > 5.9 \times 10^{33}\,$yr from an exposure of 260\,kton$\cdot$year~\cite{Super-Kamiokande:2014otb}. The next generation of experiments is being constructed: the ${\sim\,}200\,$kton Hyper-Kamiokande water Cherenkov detector and the 40\,kton liquid argon detector DUNE will probe proton lifetimes in the $10^{34}-10^{35}\,$yr regime~\cite{Hyper-Kamiokande:2018ofw,DUNE:2020fgq,Itow:2021vv}. While the projected sensitivity of these detectors in the next $10-20$\,years is impressive, it is clear that progress in proton decay searches is becoming increasingly slow and expensive, in particular because it is difficult to imagine a conventional detector significantly larger than the $\mathcal{O}(100)\,$kton scale of DUNE or Hyper-Kamiokande.

%--------------------------------------------------------------------
\section{Paleo-detectors for proton decay?} \label{sec:paleo}
%--------------------------------------------------------------------

In this work, we ask if {\it paleo-detectors} could offer a path to continuing the search for proton decay beyond Hyper-Kamiokande and DUNE. Rather than maximizing the target mass of an actively instrumented detector operated in real time, the basic paleo-detector concept is to use natural minerals formed hundreds of millions to a billion years ago as charged particle track detectors. Minerals can record and retain lattice damage (including mechanical stress, changes to the electron density, local amorphization, and isolated vacancies or interstitials) caused by the energy deposition of charged particles propagating through the crystal. If one could measure the lattice damage caused by the products of a proton decay in a one billion year old mineral, a one kg ``detector'' would suffice to match the Mton$\cdot$year exposure of DUNE and Hyper-Kamiokande. The idea of using natural minerals of geological age to search for rare events has a long history, including Refs.~\cite{Goto:1958,Goto:1963zz,Fleischer:1969mj,Fleischer:1970vm,Fleischer:1970zy,Alvarez:1970zu,Kolm:1971xb,Eberhard:1971re,Ross:1973it,Price:1983ax,Kovalik:1986zz,Price:1986ky,Guo:1988,Ghosh:1990ki,Snowden-Ifft:1995zgn,Collar:1994mj,Engel:1995gw,Collar:1995aw,Snowden-Ifft:1996dug,Jeon:1995rf,Snowden-Ifft:1997vmx,Baltz:1997dw,Collar:1999md}. Motivated by the progress of microscopy technology as well as data analysis techniques over the last decades, paleo-detectors have been proposed to search for dark matter~\cite{Baum:2018tfw,Drukier:2018pdy,Edwards:2018hcf,SinghSidhu:2019znk,Ebadi:2021cte,Acevedo:2021tbl,Baum:2021jak,Baum:2021chx,Bramante:2021dyx,Acevedo:2021tbl} and both astrophysical and atmospheric neutrinos~\cite{Baum:2019fqm,Tapia-Arellano:2021cml,Baum:2022wfc,Jordan:2020gxx}; see also Refs.~\cite{Baum:2023cct,Baum:2024eyr} for an overview of the emerging field of Mineral Detection of Neutrinos and Dark Matter.

For highly ionizing heavy nuclei (e.g., swift heavy ions, the recoiling daughters of heavy $\alpha$-decaying nuclei, or the fragments from spontaneous fission of $^{238}$U and similar isotopes) the fundamental formation mechanism and the long-term stability of lattice damage are fairly well understood. The lattice damage can be measured using a variety of imaging techniques, including electron, X-ray, and optical microscopy~\cite{Fleischer:1964,Fleischer383,Fleischer:1965yv,GUO2012233}. In the geosciences, the observed densities of the damage ``tracks'' from $\alpha$-decaying nuclei~\cite{Goegen:2000,Glasmacher:2003} and spontaneous fission fragments~\cite{Wagner:1992,Malusa:2018} in natural minerals such as apatite, mica or zircon have long been used to date minerals; fission track ages as large as ${\sim\,}0.8\,$Gyr in apatite~\cite{Murrell:2003,Murrell:2004,Hendriks:2007} and ${\sim\,}2\,$Gyr in zircon~\cite{Montario:2009} have been reported in the literature, presenting a proof-of-existence for geological environments on Earth that are sufficiently cold and stable for lattice damage from charged particles to be preserved over gigayear timescales.

The charged particles produced by a proton decaying (inside a nucleus) in a natural mineral would have much smaller maximum stopping powers ($dE/dx \sim 10^2-10^4\,$MeV/cm) than the $dE/dx \sim 10^5\,$MeV/cm regime characteristic of highly ionizing heavy nuclei. Price and Salamon observed natural ``tracks'' from Al and Si ions with a few hundred keV of nuclear recoil energy, corresponding to stopping powers in the $dE/dx \sim 10^4\,$MeV/cm regime, in 0.5\,Gyr old mica by optical microscopy following chemical etching~\cite{Price:1986}. At stopping powers of $dE/dx \sim 10^3\,$MeV/cm, Snowden-Ifft and collaborators observed damage features from $Z \sim 10$ ions (including oxygen) with few-keV nuclear recoil energies produced in controlled ion implantation and fast neutron irradiation experiments in muscovite mica using atomic force microscopy after chemical etching~\cite{SnowdenIfft:1993,Snowden-Ifft:1995rip,Snowden-Ifft:1995zgn}. 

During the last decades, various groups have demonstrated efficient readout of damage ``tracks'' produced by few-MeV $\alpha$-particles and even protons in doped sapphire (Al$_2$O$_3$:C,Mg)~\cite{Akselrod:2006,Bartz:2013,Kouwenberg:2018,Akselrod:2018,Kusumoto:2022} and in (undoped) lithium fluoride (LiF)~\cite{Bilski:2017,Bilski:2019a,Bilski:2019b,Bilski:2020} using fluorescence microscopy of {\it color center} vacancies, with effective stopping power thresholds of a few tens of MeV/cm. These results demonstrate that charged particles with stopping powers even lower than $dE/dx \sim 10^2\,$MeV/cm produce crystal damage in at least some minerals. However, comprehensive data or theoretical understanding of track formation criteria is lacking. Furthermore, while in the $dE/dx \sim 10^5\,$MeV/cm regime of highly ionizing heavy ions, the energy deposition in the crystal creates a dense (on the crystal lattice scale) string of nuclear defects, at lower stopping powers, the crystal damage seems to transition to an increasingly sparse string of isolated vacancies. Such isolated vacancies may be more mobile and susceptible to thermal annealing than dense zones of crystal damage. A dedicated experimental program is required to establish track formation criteria and the stability on geological timescales of damage features produced by charged particles in minerals in the $dE/dx \sim 10^2 - 10^3\,$MeV/cm regime.

In the absence of detailed experimental guidance on track formation and in order to make progress in exploring the potential of paleo-detectors to search for proton decay, we assume that charged particles produce lattice damage stable on geological timescales if their stopping power exceeds a threshold; we consider $(dE/dx)_{\rm min} = \{100, 500, 1000\}\,$MeV/cm thresholds, above the smallest stopping powers for which color center ``tracks'' have been observed in Al$_2$O$_3$:C,Mg and LiF. We calculate stopping powers with \texttt{SRIM/TRIM}~\cite{Ziegler:1985,Ziegler:2010}, including electronic and nuclear stopping contributions, and assume that a charged particle will give rise to a stable damage ``track'' with length
\begin{equation*}
    x = \int_0^E dE' \, \left| \frac{dE'}{dx} \right|^{-1} \times H\left( \left|\frac{dE'}{dx}\right| - \left(\frac{dE}{dx}\right)_{\rm min} \right)\,,
\end{equation*}
where $H(x)$ is the Heaviside step function and $E$ the energy with which the charged particle is emitted. In this work, we use olivine [(Mg,Fe)$_2$SiO$_4$] as an example: it is a very common mineral on both the Earth and Moon~\cite{Taylor:2009,JAUMANN201215}, forms color center vacancies~\cite{loeffler,quadery}, is stable to relatively high temperatures~\cite{demouchy}, and can be found with very low concentrations of uranium and thorium~\cite{Eggins:1998, McIntyre:2021}. Let us stress that while extremely simplified, this approach allows us to make a first step towards exploring the potential of paleo-detectors to search for proton decay. 

%--------------------------------------------------------------------
\section{Proton-decay signature} \label{sec:signal}
%--------------------------------------------------------------------

\begin{figure}
    \centering
    \includegraphics[width=\linewidth]{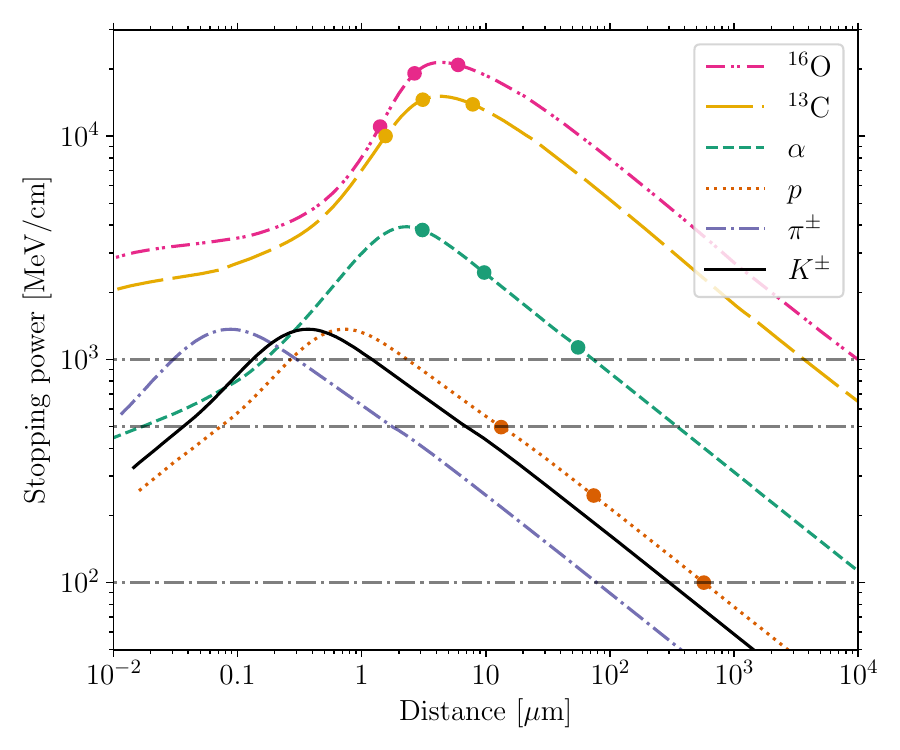}
    \caption{The stopping power ($dE/dx$) as a function of the distance ($\int_0^{E} dE' \, |dE'/dx|^{-1}$) of various charged particles (see legend) in olivine. The dots on the $^{16}$O, $^{13}$C, $\alpha$-particle and proton ($p$) lines mark where an ion with (left to right) $1,\,3,\,10\,$MeV kinetic energy would ``start.'' Note that $\pi^\pm$ and $K^\pm$ with energies relevant to us (typical $\pi^\pm$ produced in DIS of neutrinos have energies ${\sim\,}10^2-10^3$\,MeV; the $K^+$ from $p$-decay have $\mathcal{O}(100)\,$MeV energy) start at much larger distances/much smaller stopping powers than what is shown on this plot. The horizontal dash-dotted gray lines indicate stopping powers of $dE/dx = \{100, 500, 1000\}$\,MeV/cm.}
    \label{kaon_stopping_power}
\end{figure}

Typical decays of protons arising in BSM theories are into a lepton and a light meson, e.g., $p \to e^+ \pi^0$ or $p \to \bar{\nu} K^+$. In the rest frame of the proton, the lepton and the light meson have momenta of a few hundred MeV. Positrons and photons arising from $\pi^0 \to 2\gamma$ decays have too small stopping powers to leave permanent lattice damage in minerals. Charged kaons, on the other hand, might give rise to measurable signatures in paleo-detectors: a $K^+$ with a kinetic energy of $T_{K^+} \sim 100\,$MeV has a range of a few cm in olivine. Towards the end of its trajectory, its stopping power exceeds 100\,MeV/cm; see Fig.~\ref{kaon_stopping_power}, where we show the stopping power for charged kaons in olivine as a function of distance from the endpoint together with various other charged particles that are relevant to our discussion. For track formation thresholds $\sim 10^2-10^3\,$MeV/cm, a kaon from $p \to \bar{\nu} K^+$ would first propagate a few cm through olivine without producing permanent lattice damage. However, as the kaon approaches the Bragg peak, its stopping power becomes sufficiently large to produce what we will refer to as a ``kaon endpoint'' track. 

In olivine and most other minerals, protons are not free but are bound in nuclei. Thus, the decaying proton will not be at rest relative to the mineral, but will have a typical momentum given by the Fermi momentum of protons in the nucleus (for oxygen, $p_F \sim 200\,$MeV). In addition, final-state interactions of the kaon with the nucleus lead to an additional broadening of the kaon spectrum. For example, modeling $p \to \bar{\nu}K^+$ decays in $^{16}$O with \texttt{FLUKA2021}~\cite{Bohlen:2014buj,Ferrari:2005zk,fluka_website}, we find a mean energy of $\overline{T}_{K^+} = 80\,$MeV and an RMS smearing of $T^{\rm rms}_{K^+} = 40\,$MeV. The nuclear remnant (e.g., $^{15}$N for proton decay in $^{16}$O) typically stays intact but recoils with an energy $T_N \sim p_F^2/2m_N \sim 1\,$MeV. Due to its relatively large charge, the stopping power of such a nucleus is $dE/dx \sim 10^4\,$MeV/cm (see Fig.~\ref{kaon_stopping_power}), much larger than any stopping power threshold we consider here; the associated track has a typical length of ${\sim}\,2\,\mu$m. Note that this length would be almost unchanged for $p \to e^+ \pi^0$ decays since it is controlled by the momentum of the decaying proton with small corrections from final state interactions.

\begin{figure}
    \begin{centering}
    \includegraphics[width=\linewidth]{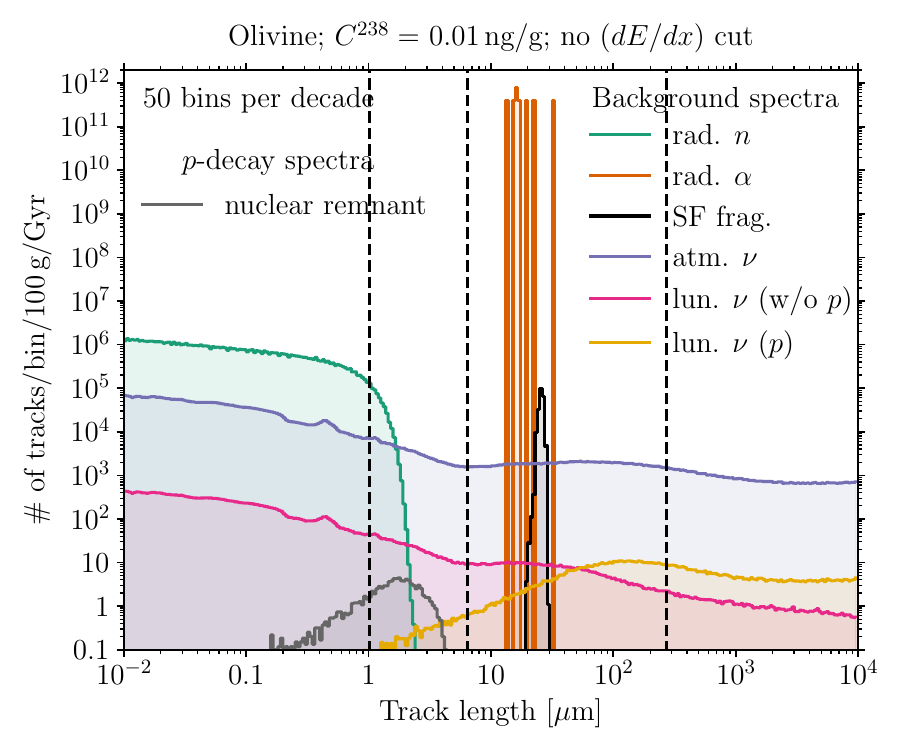}
    \caption{Track length spectra in olivine normalized to an exposure of 100\,g$\cdot$Gyr, using 50 logarithmically spaced bins per decade. The various radiogenic backgrounds (``rad. $n$'': radiogenic neutron induced nuclear recoils, ``rad. $\alpha$'': $\alpha$-particle tracks, ``SF frag.'': spontaneous fission fragments) are normalized to a $^{238}$U concentration of $10^{-11}$ per weight, or $0.01\,$ng/g. The ``atm. $\nu$'' spectrum shows the background induced by the interactions of atmospheric neutrinos on Earth. The ``lun. $\nu$'' spectra show the background induced on the Moon by the interactions of cosmic-ray produced neutrinos; note that we have divided this background into the spectra produced by protons and all other charged particles. The ``$p$-decay remnant'' distribution shows the track length spectrum produced by the $^{15}$N nuclear remnant of proton decay (for simplicity, we modeled only decays in $^{16}$O); for illustration this spectrum is normalized to 100\,events/100\,g$\cdot$Gyr, which would correspond to a lifetime of order $\tau_p \sim 3 \times 10^{32}\,$yr, approximately an order of magnitude below the current Super-Kamiokande limit. All of these spectra are shown without applying any stopping power cuts. The black vertical dashed lines indicate the length of the kaon endpoint tracks for (right to left) stopping power thresholds of $(dE/dx)_{\rm min} = \{100, 500, 1000\}\,$MeV/cm.
    } 
    \label{track_length_comparison}
    \end{centering}
\end{figure}

Let us summarize the signatures of proton-decay in paleo-detectors. For the $p \to e^+ \pi^0$ decay mode, the only potential signature is the track produced by the ${\sim}\,1\,$MeV nuclear remnant. Given the fairly large stopping powers ($dE/dx \sim 10^4\,$MeV/cm) of these remnants, there is little reason to doubt that they will produce lattice damage stable on geological timescales in natural minerals. However, for proton lifetimes larger than current limits from Super-Kamiokande, any tracks from the nuclear remnants of proton decay will be buried below an overwhelming background induced by radiogenic neutrons and atmospheric neutrinos which we will discuss below (see also Fig.~\ref{track_length_comparison}). 

For the $p \to \bar{\nu} K^+$ decay mode the situation is more hopeful: if charged particles with stopping powers smaller than ${\sim}\,10^3\,$MeV/cm give rise to lattice damage that is stable on geological timescales, a proton decay will result in both a nuclear remnant track and a ``kaon endpoint'' track. The kaon endpoint will be separated from the nuclear remnant by a few cm, making any association of the endpoint with the nuclear remnant track extremely difficult in the presence of background nuclear recoils. However, the kaon endpoint track alone could suffice as a signature, allowing one to search for proton decay with paleo-detectors. In Fig.~\ref{track_length_comparison}, we plot both the track length spectrum of the $^{15}$N remnant nuclei associated with proton decay in $^{16}$O and the characteristic lengths of kaon endpoint tracks for several stopping power thresholds in olivine.

%--------------------------------------------------------------------
\section{Backgrounds, on the Moon?} \label{sec:bkg}
%--------------------------------------------------------------------

The most important sources of backgrounds for a proton decay search with paleo-detectors are natural radioactivity and the interactions of cosmogenic particles, in particular, cosmic ray produced neutrinos and muons. The relative natural abundances and lifetimes of heavy radioactive elements render $^{238}$U the most relevant source of radiogenic backgrounds in paleo-detectors. Natural olivine samples occur with a wide range of uranium concentrations, ranging from tens of parts per billion to samples with such low concentration that only upper limits of $<30\,$parts per trillion (ppt) could be measured~\cite{Eggins:1998,McIntyre:2021}. For all results in this work we assume a $^{238}$U concentration of $0.01\,$ng/g (10\,ppt per weight), although uranium concentrations even orders of magnitudes larger would not change our conclusions.

The decays of $^{238}$U and its daughters lead to various backgrounds. First, the eight $\alpha$-decays in the uranium decay chain each produce 
\begin{equation*}
    n_\alpha \sim \frac{4 \times 10^{11}}{100\,{\rm g}\cdot{\rm Gyr}} \times \left( \frac{C^{238}}{0.01\,{\rm ng/g}}\right)
\end{equation*}
$\alpha$-particles with energies between $4.2\,$MeV and $7.7\,$MeV. As long as charged particles with stopping powers larger than ${\sim}\,10^3\,$MeV/cm give rise to permanent lattice damage, these $\alpha$-particles will produce ``tracks'' along their entire trajectory with a typical length of a few tens of microns, see Figs.~\ref{kaon_stopping_power} and~\ref{track_length_comparison}. Each $\alpha$-decay also gives rise to a $\mathcal{O}(10)\,$keV recoil of the heavy nuclear remnant. The $Z \sim 90$ nuclear remnants have large stopping powers, producing tens of nm long ``$\alpha$-recoil tracks.'' While the number of $\alpha$-particle and $\alpha$-recoil tracks may seem overwhelmingly large, they can be distinguished from the kaon endpoint signal from proton decay by both their characteristic track length and ``clustering.'' The summed half-life of all decays in the uranium chain following the first $^{238}{\rm U} \to {^{234}{\rm Th}}+\alpha$ step is $0.3\,$Myr. In a billion year old paleo-detector, the characteristic signature from the uranium chain is thus not isolated pairs of $\alpha$-particle and $\alpha$-recoil tracks, but clustered patterns of eight $\mathcal{O}(10)\,$nm long high-$dE/dx$ $\alpha$-recoil tracks with eight $\mathcal{O}(10)\,{\mu}$m low-$dE/dx$ $\alpha$-particle tracks attached.

The nuclei in the uranium decay chain, in particular $^{238}$U, also undergo spontaneous fission. The $Z\sim50$ fission fragments have energies of ${\sim}\,50 - 100\,$MeV and large $dE/dx$, giving rise to $\mathcal{O}(10)\,\mu$m long fission fragment tracks. We use \texttt{FREYA}~\cite{Verbeke:2015hka} to model the distribution of fission fragments. Since the fission fragments are emitted back-to-back, we treat the tracks from the two daughters as a single longer track, shown as the ``SF frag.'' histogram in Fig.~\ref{track_length_comparison}. Each spontaneous fission event also emits $\sim\,$2 fast neutrons, which in turn give rise to nuclear recoils when scattering off the nuclei making up the paleo-detector. In addition, $(\alpha,n)$-reactions also produce fast neutrons. We use a custom version of \texttt{SOURCES-4a}~\cite{sources4a:1999} to model the neutron spectrum from spontaneous fission and $(\alpha,n)$-reactions of $\alpha$-particles from the uranium chain in olivine. 

We employ our own Monte Carlo simulation to calculate the neutron-induced nuclear recoil spectra using \texttt{TENDL-2017} neutron-nucleus cross sections~\cite{Koning:2012zqy,Soppera:2014zsj,Sublet:2015,Fleming:2015,Rochman:2016}. The tracks produced by the interactions of radiogenic neutrons are shorter than $3\,\mu$m in olivine (see the ``rad.~$n$'' spectrum in Fig.~\ref{track_length_comparison}). The large number of these tracks seems to preclude the possibility of using the ${\sim}\,2\,\mu$m nuclear remnant tracks as a signal to search for proton decay in paleo-detectors. Comparing the radiogenic neutron induced background to the kaon endpoint proton-decay signature, the stopping power threshold over which charged particles produce permanent lattice damage plays a crucial role. For the range we consider here, $(dE/dx)_{\rm min} = 10^2-10^3\,$MeV/cm, the radiogenic neutron induced background is practically unaltered. However, for a threshold of $10^3\,$MeV/cm, kaon endpoints are only ${\sim}\,1\,\mu$m long, burying them under the radiogenic neutron background. For $\lesssim 500\,$MeV/cm thresholds, on the other hand, kaon endpoints are $\gtrsim 6\,\mu$m long, a factor of a few longer than the tracks that could be induced from the most energetic radiogenic fast-neutron induced nuclear recoils.

\begin{figure}
    \begin{centering}
    \includegraphics[width=\linewidth]{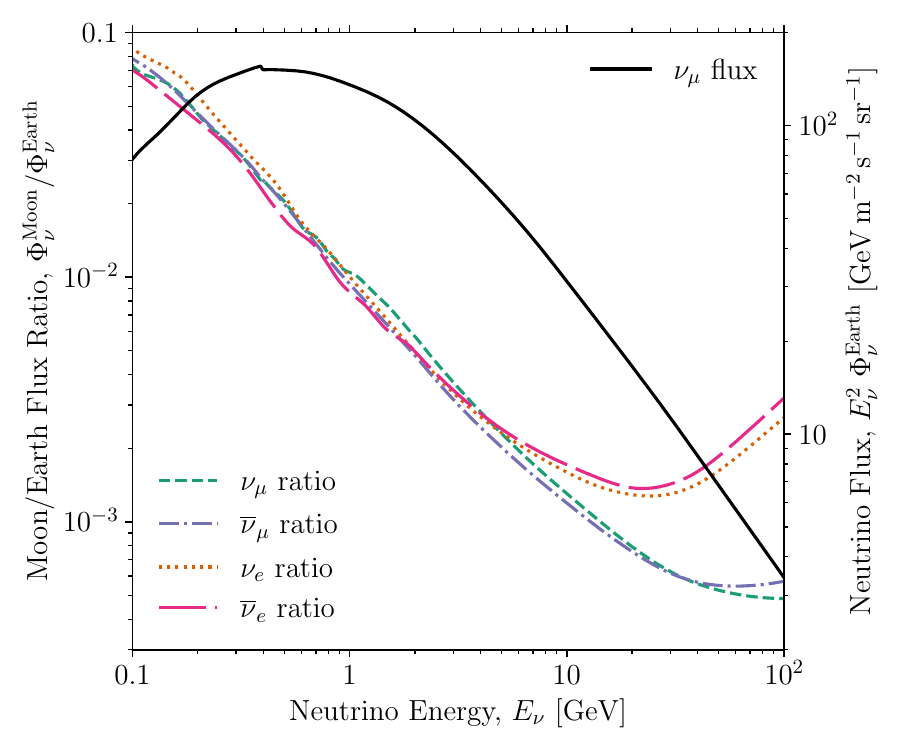}
    \caption{The colored lines (see left legend and left $y$-axis) show the {\it ratio} of the flux of neutrinos produced by cosmic ray interactions on the Moon to the atmospheric neutrino flux on Earth for each neutrino flavor. For reference, the black solid line (right $y$-axis) shows the atmospheric $\nu_\mu$ flux on Earth. Note that neutrino-induced kaon production turns on around $E_\nu \sim 1-2\,$GeV.}
    \label{figure_flux_ratio}
    \end{centering}
\end{figure}

Let us turn to cosmogenic backgrounds. Neutrinos emitted by our sun and supernovae cause nuclear recoils in a paleo-detector predominantly via coherent elastic scattering off the nuclei making up the mineral. As the corresponding recoil tracks are either orders of magnitude shorter in length or orders of magnitude less in number compared to tracks induced by radiogenic neutrons~\cite{Baum:2019fqm,Tapia-Arellano:2021cml}, we omit these neutrino background contributions in Fig.~\ref{track_length_comparison}.
Above neutrino energies of ${\sim}\,0.1\,$GeV, atmospheric neutrinos, produced by the interactions of cosmic rays in the atmosphere, dominate the neutrino contributions on Earth; in Fig.~\ref{figure_flux_ratio}, the black line shows the atmospheric $\nu_\mu$ flux~\cite{Battistoni:2005pd,Honda:2015fha}. These neutrinos are sufficiently energetic to interact with the nuclei making up olivine (or any other paleo-detector) via quasi-elastic, resonant, and (deep) inelastic scattering, which can produce not only a recoiling nuclear remnant but also lighter nuclear fragments and/or particles that can both directly give rise to lattice damage and cause secondary nuclear recoils. To estimate the associated background, we model atmospheric $\nu_\mu$, $\overline{\nu}_\mu$, $\nu_e$, and $\overline{\nu}_e$ interactions with \texttt{FLUKA2021}~\cite{Ferrari:2005zk,Bohlen:2014buj,fluka_website}, including secondary interactions of leptons, mesons, baryons, and nuclear fragments produced in the primary neutrino-nucleus interaction. 

In Fig.~\ref{track_length_comparison}, the associated track length spectrum is labeled ``atm. $\nu$''. We can see that atmospheric neutrinos contribute the most relevant background for the ($\lesssim 500\,$MeV/cm threshold) kaon endpoint signal from $p \to \bar{\nu} K^+$ proton decays in paleo-detectors. While not explicitly shown in Fig.~\ref{track_length_comparison}, a particularly worrying background is charged kaons created by atmospheric neutrinos. For all other backgrounds one could, at least in principle, imagine a substantial suppression if the microscopic readout of the paleo-detector allowed one to reconstruct not only the length of a track (used as a discriminating variable in Fig.~\ref{track_length_comparison}), but a quantity that would allow for particle identification, e.g., the stopping power profile along the track. This may be possible when using, for example, fluorescent microscopy sensitive to optically active vacancies created by a charged particle: the density of color centers along a ``track'' would be (non-linearly) related to the stopping power. 

The current upper limit from Super-Kamiokande~\cite{Super-Kamiokande:2014otb}, $\tau(p \to \bar{\nu} K^+) > 5.9 \times 10^{33}\,$yr corresponds to $\lesssim 6$ kaons per 100\,g per Gyr from proton decays. In an olivine sample on Earth, atmospheric neutrino primary and secondary interactions would create ${\sim}\,400$ kaon endpoints per 100\,g per Gyr. While one may imagine reducing this background by associating kaon endpoints with either the nuclear remnant track from proton decay or the cluster of tracks produced in an inelastic atmospheric neutrino interaction, such association does not seem feasible given that a charged kaon first travels a few cm through the mineral, with stopping power below any of the $dE/dx$ thresholds considered here, before producing an endpoint track (see Fig.~\ref{kaon_stopping_power}). 

Intriguingly, humanity is currently developing the means to return to and expand our presence on the Moon~\cite{artemis}, with a variety of proposals already put forth for astroparticle and high energy physics experiments there~\cite{Silk:2020bsr,Beacham:2021lgt,artemissci,Gaspert:2023ezv}. The Moon does not possess an atmosphere, thus, cosmic rays interact directly with the Moon's regolith. Due to the regolith's density, most of the cosmic ray interaction products whose decays subsequently produce neutrinos (most importantly, charged pions), stop before they decay. Consequently, the energetic cosmic-ray induced neutrino flux on the Moon is suppressed compared to Earth~\cite{Miller:2006cn,Li:2017oec,Demidov:2020bff}. 

In Fig.~\ref{figure_flux_ratio}, we show the ratio of the cosmic-ray produced neutrino flux on the Moon to the atmospheric neutrino flux on Earth. For neutrino energies $\lesssim 2 \, $GeV, we use the ``conventional'' lunar neutrino fluxes calculated for the various neutrino flavors in Ref.~\cite{Demidov:2020bff}. At higher neutrino energies, we simply extrapolate from the low energy part of the conventional spectrum with the power law calculated in Ref.~\cite{Miller:2006cn} for all flavors. At neutrino energies $\gtrsim 10 \,$GeV, we see the suppression of the lunar neutrino fluxes relative to the atmospheric neutrino fluxes begin to reverse due to the contribution of the ``prompt'' neutrino flux. 

Rather than being produced by secondary pions and kaons, prompt neutrinos are produced in the much faster decays of heavy-flavor hadrons. While the prompt neutrino flux is smaller on the Moon than on Earth, the suppression is much milder than for the conventional flux because heavy-flavor hadrons have much shorter lifetimes than light mesons and hence tend to decay in flight even in the dense environment of the Moon's regolith. Thus, we make the conservative assumption that the prompt neutrino flux does not differ between the Earth and the Moon. We take the prompt fluxes calculated for the Earth's atmosphere from Ref.~\cite{Fedynitch:2015zma} and add them to the conventional fluxes described above.

The primary and secondary interactions of ``lunar neutrinos'' give rise to a broad distribution of background tracks in lunar olivine. In the most relevant neutrino energy range, ${\sim\,}1-100\,$GeV, the neutrino flux on the Moon is two to three orders of magnitude smaller than on Earth. Correspondingly, we can see in Fig.~\ref{track_length_comparison} that the ``lun. $\nu$'' background is a few orders of magnitude smaller than the atmospheric neutrino background on Earth -- note that we show the proton induced background spectrum separately since protons are the charged particles most similar to charged kaons in stopping power (see Fig.~\ref{kaon_stopping_power}) and would thus be the most challenging ones to differentiate from kaons with a $dE/dx$-sensitive readout technique. Charged pions are less of a concern due to their smaller relative rate compared to protons.

While on Earth, we found a rate of ${\sim}\,400$ kaon endpoints per 100\,g per Gyr in olivine, we instead expect $\sim 0.5$ kaons per 100\,g per Gyr in lunar olivine. In general, the neutrino-induced kaon production cross section is low in the few-GeV region most applicable for the cosmic-ray produced neutrino flux due to both the mass of the kaon and because the relevant resonances are small~\cite{Formaggio:2012cpf}. The cross sections of the relevant neutrino processes, led by the charged current interactions $\nu_\mu n \rightarrow \mu^- K^+ \Lambda^0$ and $\nu_\mu p \rightarrow \mu^- K^+ p$, are poorly studied, with the leading neutrino event generator~\cite{Ferrari:2005zk,Bohlen:2014buj,Andreopoulos:2009rq,Lalakulich:2011eh,Golan:2012wx,Golan:2012rfa} predictions varying substantially and only one published modern measurement~\cite{MINERvA:2016zyp}. Notably, however, this measurement is somewhat relevant to our event rate prediction uncertainty since it is based on similar, few-GeV $\nu_\mu$ interactions [on plastic scintillator (CH)]; 20-30\%-level experimental uncertainties and an overall 15\% disagreement with generator predictions are reported.

The interactions of cosmic rays with Earth's atmosphere or the lunar regolith produce not only neutrinos but also muons. Muon interactions in the vicinity of a paleo-detector can produce energetic neutrons and a variety of spallation fragments, including charged kaons. Unlike neutrinos, the flux of muons is attenuated approximately exponentially with depth, although their mean energy also changes with the overburden~\cite{Kudryavtsev:2003aua,Mei:2005gm}. A dedicated chain of simulations modeling the cosmogenic muon spectra at the surface, propagating them through the overburden, and simulating their interactions in and near a paleo-detector would be required to fully estimate the cosmogenic muon induced background. Such a simulation is beyond the scope of this work; we will only make a simple estimate here to show that the cosmogenic muon background can be mitigated. 

While on Earth's surface, the cosmogenic muon flux at energies $\lesssim 100 \,$TeV is dominated by the conventional component, at depth the prompt contribution is increasingly important. This is because the prompt muon energy spectrum is harder than the conventional muon spectrum, and thus attenuated more slowly with depth. We use the results of Ref.~\cite{Fedynitch:2015zma} for the conventional and prompt muon fluxes on Earth, and then approximate the suppression of the conventional muon fluxes on the Moon as that of the conventional muon neutrino fluxes shown in Fig.~\ref{figure_flux_ratio}. As with the prompt neutrino fluxes, we assume the prompt muon flux is the same on the Earth and the Moon.

With the conventional and prompt muon fluxes on the Moon described above, we calculate the attenuation at depth following Ref.~\cite{Lipari:1991ut}. At a depth of $5\,$km rock, we find an integrated muon flux of $\Phi_\mu \sim 10^3\,{\rm cm}^{-2}\,{\rm Gyr}^{-1}$ and a mean muon energy of $\overline{E}_\mu \sim 700\,$GeV. We stress that this is almost entirely due to the prompt flux -- if we had included only the conventional muon flux in our estimate, we would have found $\Phi_\mu^{\rm conv} \sim 0.2\,{\rm cm}^{-2}\,{\rm Gyr}^{-1}$ at 5\,km depth on the Moon. In the depth regime dominated by the prompt flux, the integrated muon flux is reduced roughly by one order of magnitude per km rock overburden; for example, at 6\,km overburden we find $\Phi_\mu \sim 10^2\,{\rm cm}^{-2}\,{\rm Gyr}^{-1}$ and $\overline{E}_\mu \sim 600\,$GeV. Even the extremely conservative requirement that the total muon flux should be $\ll 10^{-2}\,{\rm cm}^2\,{\rm Gyr}^{-1}$, i.e., less than 1\,muon passing through a kg-size paleo-detector per Gyr, would be satisfied for an overburden of $\gtrsim 10\,$km rock. Paleo-detectors at such depths on the Moon could be obtained, for example, from the cores of deep boreholes similar to those on Earth~\cite{KREMENETSKY198611,Hirschmann1997,Blattlereaar2687}.

The corresponding muon-induced fast neutron flux at ${\sim}\,5$\,km depth on the Moon is ${\sim}\,10^2\,{\rm cm}^{-2}\,{\rm Gyr}^{-1}$, with the normalization scaling roughly equivalent to that of the muon fluxes with depth. The neutron interactions at ${\sim}\,5$\,km will thus typically induce recoils of the constituent nuclei in olivine with a rate less than that of the lunar neutrino flux, becoming even more suppressed at greater depths. The more concerning background for the proton decay kaon endpoint signal is the production of charged kaons in muon spallation. The integrated muon-proton deep inelastic scattering (DIS) cross section in the relevant energy range is ${\sim\,}\,10\,\mu$b~\cite{Dutta:2000hh}, and from the results of the COMPASS experiment we estimate the multiplicity of charged kaons produced in DIS to be ${\sim}\,0.15$~\cite{Stolarski:2023}. 

Given the muon flux, muon-proton cross section and charged kaon multiplicity, we expect that cosmogenic muons give rise to ${\sim}\,0.1$ kaons per 100\,g per Gyr in an olivine paleo-detector at a depth of 5\,km. This charged kaon background is again less than what is expected from lunar neutrinos and further suppressed approximately as the muon flux with depth. We leave a more precise determination of the required depth for paleo-detector retrieval for a lunar proton decay search to future work; let us here only comment that launch vehicles for future lunar missions currently under development such as NASA's SLS have envisaged payloads of ${\sim}\,100\,$tons~\cite{sls,artemissci}, making it possible to imagine delivering a rig able to drill many km deep to the Moon.

%--------------------------------------------------------------------
\section{Discussion and Summary} \label{sec:discussion}
%--------------------------------------------------------------------

Above, we have discussed the signatures of proton-decay in paleo-detectors and the most relevant sources of backgrounds. The characteristic signature of a $p \to \bar{\nu} K^+$ proton-decay event in a paleo-detector will be a few-micron long track from the nuclear remnant and a kaon endpoint track (with length dependent on the stopping power threshold). Since the kaon endpoint is separated from the nuclear remnant track by a few cm, making any association of the endpoint with the nuclear remnant track appears extremely difficult. Regardless of the stopping power threshold, the nuclear remnant tracks are buried under a large background induced by radiogenic neutrons as well as cosmogenic neutrinos; thus, we focus on the kaon endpoint track as the sole signal of proton decay here.

An irreducible background to proton-decay induced kaon endpoints arises from the interactions of {\it atmospheric} neutrinos on Earth, which lead to ${\sim}\,400$ kaon endpoints per 100\,g per Gyr in olivine. This rate can be compared to the current upper limit from Super-Kamiokande, $\tau(p \to \bar{\nu} K^+) > 5.9 \times 10^{33}\,$yr, which corresponds to $\lesssim 6\,$ kaons per 100\,g per Gyr from proton decay. Intriguingly, the Moon offers a reprieve from this irreducible background. Since the Moon does not have an atmosphere, cosmic rays interact directly with the regolith on the Moon, resulting in a spectrum of cosmic ray induced neutrinos much less energetic than atmospheric neutrinos on Earth. We estimate that the number of kaons produced by cosmogenic neutrinos on the Moon is ${\sim}\,0.5$ per 100\,g per Gyr. Another relevant source of kaons are the interactions of cosmic ray induced muons; we estimate that both on the Moon and on Earth, paleo-detector samples would have to be obtained from a depth of ${\sim\,}5\,$km rock to suppress the rate of cosmogenic muon produced kaons to ${\sim}\,0.1$ per 100\,g per Gyr. Thus, irreducible backgrounds are sufficiently low to allow for a proton-decay search beyond the reach of Super-Kamiokande and even Hyper-Kamiokande/DUNE with lunar paleo-detector samples obtained from a depth larger than 5\,km.

While the damage features in minerals extracted from the Moon are less well studied than in terrestrial samples, prior analyses of lattice damage due to cosmic rays incident on samples near the lunar surface and more general investigations of lunar geology are promising. Damage tracks from highly ionizing iron-group nuclei associated with cosmic rays have been measured in mineral samples from the lunar regolith using optical microscopy techniques after chemical etching~\cite{Crozaz:Moon,Fleischer:Moon,Bhandari:Moon,Plieninger:Moon}. Aside from the erasure of tracks caused by impact deformation of samples acquired near the lunar surface~\cite{Fleischer:Shock73,Fleischer:Shock74}, the tracks formed by these ions with stopping powers of $dE/dx \sim 10^5\,$MeV/cm have been shown to persist over geological timescales. 

Although a detailed study taking into account the geological conditions on the Moon will be necessary, let us stress that due to the relatively low temperature gradients and minimal geological activity typical of the lunar crust at depths of $\gtrsim 2\,$km over the last ${\sim\,}\,$Gyr~\cite{Taylor:2009,JAUMANN201215}, the geological environment on the Moon appears quite favorable for the formation and retention of lattice damage from low-$dE/dx$ particles in paleo-detectors at depths relevant for proton decay searches. As an aside, these geological conditions may make lunar samples an interesting alternative to samples from the Earth not only for proton-decay searches, but also for paleo-detector searches for dark matter or astrophysical neutrinos proposed in Refs.~\cite{Baum:2018tfw,Drukier:2018pdy,Edwards:2018hcf,SinghSidhu:2019znk,Ebadi:2021cte,Acevedo:2021tbl,Baum:2021jak,Baum:2021chx,Bramante:2021dyx,Acevedo:2021tbl,Baum:2019fqm,Tapia-Arellano:2021cml,Baum:2022wfc,Jordan:2020gxx}.

\begin{figure*}
    \centering
    \includegraphics[width=0.49\linewidth]{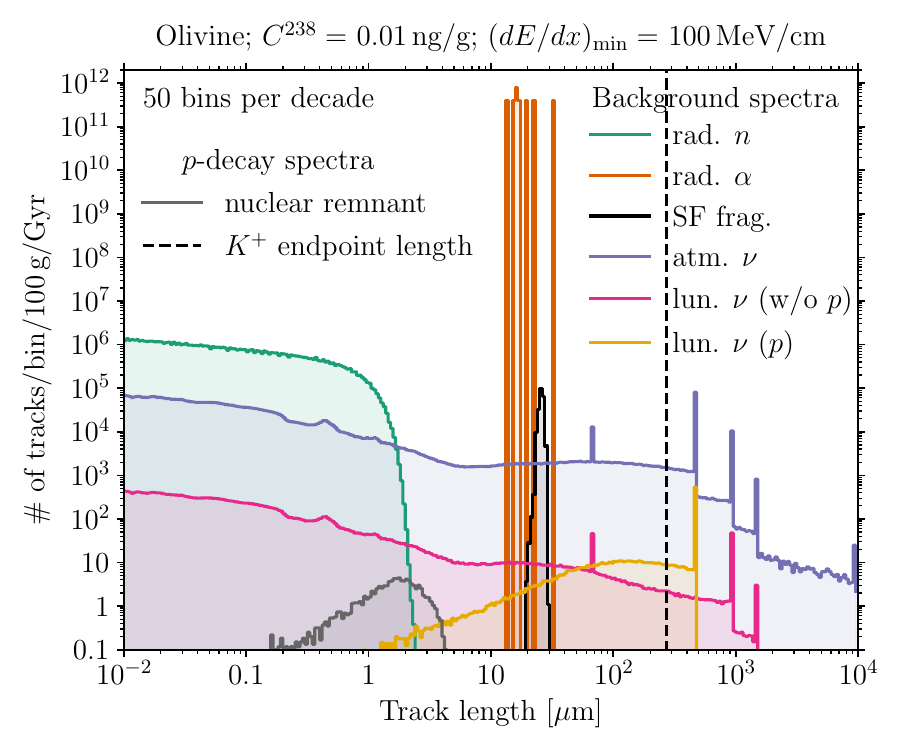}
    \includegraphics[width=0.49\linewidth]{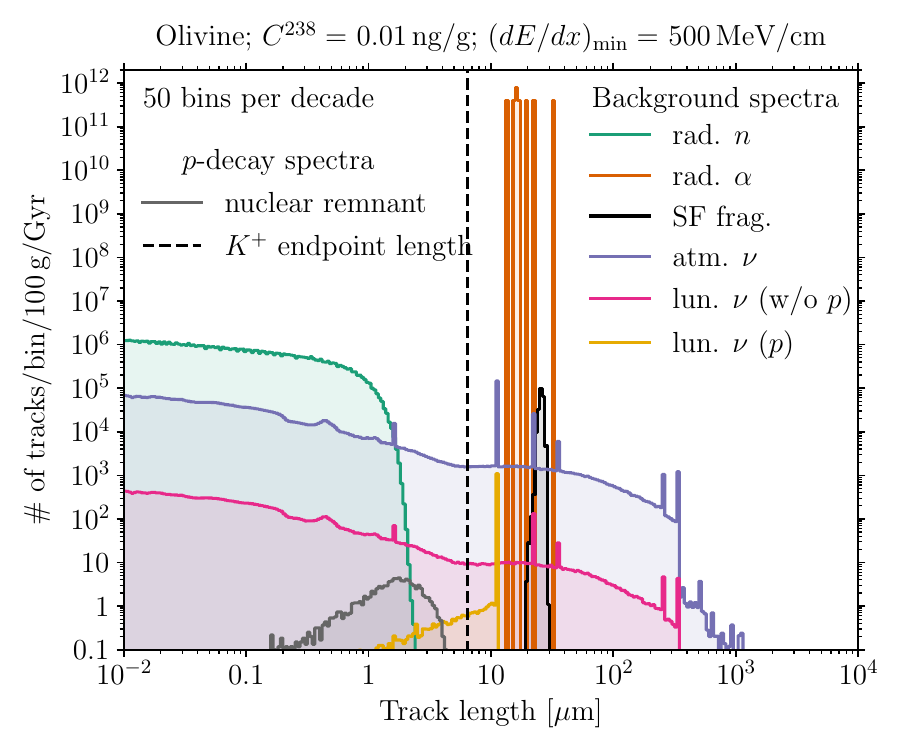}
    \caption{Same as Fig.~\ref{track_length_comparison}, but including ``track formation'' stopping power thresholds of $\left( dE/dx\right)_{\rm min} = 100\,$MeV/cm (left) and $500\,$MeV/cm (right).}  
\label{kaontrack_and_neutrino_background}
\end{figure*}

In Fig.~\ref{kaontrack_and_neutrino_background}, we show the reducible backgrounds for a $p \to \bar{\nu} K^+$ proton-decay search with a lunar olivine paleo-detector taken from sufficiently large depth so that we can ignore the cosmogenic muon induced background. In the left panel, we show the results for an $(dE/dx)_{\rm min} = 100\,$MeV/cm effective stopping power threshold for the formation of lattice damage that is stable on geological timescales. In this case, the kaon endpoint signals would have a characteristic track length of ${\sim}\,270\,\mu$m. In Fig.~\ref{kaontrack_and_neutrino_background}, we use 50 bins per decade in track length to show the spectra, roughly corresponding to a read-out resolution of $2\,\%$ in track length. In the vicinity of the ${\sim}\,270\,\mu$m signal, we find a background of ${\sim\,}10$\,tracks/bin/100\,g/Gyr from cosmogenic neutrino induced protons on the Moon, and ${\sim\,}2$\,tracks/bin/100\,g/Gyr from higher-$Z$ nuclei produced by cosmogenic neutrino interactions. Since these particles have different stopping powers than charged kaons (see Fig.~\ref{kaon_stopping_power}) these backgrounds can be suppressed if the lattice damage is measured with a technique sensitive to $dE/dx$, allowing for particle identification. As mentioned above, one such possibility could be fluorescent microscopy, which has been shown to be sensitive to lattice damage produced by even lower stopping powers and where, in principle, the density of color centers along the tracks can be used to deduce the stopping power profile of the particle producing the lattice damage. Given that, as shown in Fig.~\ref{kaon_stopping_power}, the stopping power of protons is quite similar to that of charged kaons, it is difficult to estimate how effective such a background suppression can be without a dedicated experimental program. From our background estimates, track discrimination between protons and kaons leading to an effective suppression of proton background tracks by a factor of $>20$ would reduce this background below the irreducible kaon background on the Moon, while a suppression by a factor of a few would suffice for a proton decay search with sensitivity better than the current limit from Super-Kamiokande.

In the right panel of Fig.~\ref{kaontrack_and_neutrino_background}, we show the reducible backgrounds for $(dE/dx)_{\rm min} = 500\,$MeV/cm, for which the kaon endpoint signals would have a characteristic length of ${\sim}\,6\,\mu$m. The dominant background at these track lengths is again due to cosmogenic neutrinos, but the balance between tracks from protons and from higher-$Z$ nuclei is now inverted: using again 50 bins per decade in track length, we find ${\sim\,}10$\,tracks/bin/100\,g/Gyr from higher-$Z$ nuclei, and ${\sim\,}0.5$\,tracks/bin/100\,g/Gyr from cosmogenic neutrino induced protons. As we argued above, discriminating background tracks from the kaon endpoint signal tracks via a $dE/dx$-sensitive readout technique is much easier for higher-$Z$ tracks than for proton tracks. On the other hand, we can see in Fig.~\ref{kaontrack_and_neutrino_background} that for $(dE/dx)_{\rm min} = 500\,$MeV/cm, the kaon endpoint tracks are shorter than the ${\sim}\,10^{12}$ $\alpha$-particle tracks per 100\,g per Gyr we expect in an olivine sample with 0.01\,ng/g uranium. At first sight, these tracks seem to constitute an overwhelmingly large background: if only one in a trillion of these tracks has a measured track length a factor of $\sim$\,2 shorter than expected, it would appear similar to a charged kaon track (up to the different $dE/dx$). However, the $\alpha$-tracks do appear clustered with connected patterns of eight $\mathcal{O}(10)\,$nm long high-$dE/dx$ $\alpha$-recoil tracks with eight $\mathcal{O}(10)\,{\mu}$m low-$dE/dx$ $\alpha$-particle tracks attached. This characteristic pattern gives reasons to hope for extremely efficient rejection of $\alpha$ backgrounds relevant for isolated kaon endpoint signals in the $(dE/dx)_{\rm min} = 500\,$MeV/cm scenario.
 
%--------------------------------------------------------------------
\section{Conclusions} \label{sec:conclusions}
%--------------------------------------------------------------------
In this work, we have discussed the possibility of using paleo-detectors to search for proton decay. If one measures the lattice damage caused by the products of a proton decay in a one billion year old mineral, a one kg ``detector'' would suffice to match the future expected Mton$\cdot$year exposure of DUNE and Hyper-Kamiokande. This idea clearly is speculative and presents significant technical challenges. In principle, the signature of proton decay in a paleo-detector could be rich: a few-micron-long, high-$dE/dx$ track from the nuclear remnant and additional lattice damage from the charged proton decay product(s). In light of radiogenic and cosmogenic backgrounds, however, the lattice damage that might be produced by charged kaons from $p \to \bar{\nu} K^+$ decays at the end of the kaon's range appears to be the only feasible signature. Sufficient suppression of the irreducible background of charged kaons produced by atmospheric neutrinos in terrestrial paleo-detector samples requires that such an experiment must be conducted using samples from the Moon. For the charged kaon background arising from cosmogenic muons to be suppressed, lunar paleo-detector samples must be extracted from depths greater than 5\,km. Given the current push to return to the Moon, sourcing minerals from such depths seems less far-fetched than a first impression would suggest.

In addition to the challenges associated with background mitigation, numerous other relevant questions remain. Here, we simply assumed that charged particles with stopping powers above certain thresholds give rise to lattice damage that is stable on geological timescales. Multiple aspects of lattice damage production in the $10^2-10^3\,$MeV/cm stopping power regime remain poorly understood, however. It is likely that such damage consists of relatively sparse (on the crystal lattice scale) isolated vacancies. In order to gain a deeper understanding of the thermal annealing and mobility of such vacancies on geological timescales, a dedicated experimental program must be conducted. A microscopy technique that is capable of efficiently reading out these damage features with particle discrimination in macroscopic volumes is required. Minerals which are both common on the Moon and susceptible to the formation of color center defects that can be read out using fluorescent microscopy techniques, e.g. olivine, appear to be the most promising for paleo-detectors.

The search for proton decay has yielded remarkable results over the past 50 years, with the next generation of experiments, Hyper-Kamiokande and DUNE, exploring the $\tau_p \sim 10^{34}-10^{35}\,$yr regime. However, if the proton lifetime is beyond the reach of these experiments, the field is in urgent need of new ideas. The path of building ever larger actively instrumented detectors with sufficient mitigation of correspondingly larger backgrounds is reaching a point beyond which it seems unfeasible to proceed at any reasonable cost. We hope that this work will open a window to exploring new ideas that will push the search for proton decay to the final frontier. 

%--------------------------------------------------------------------
\section*{Acknowledgements}
We thank Juan I.~Collar, Katherine Freese and Johnathon Jordan for many useful discussions. 
SB acknowledges support from the DFG under grant 396021762 - TRR 257: Particle Physics Phenomenology after the Higgs Discovery. The work of SB was in part performed at the Aspen
Center for Physics, which is supported by National Science Foundation grant PHY-2210452. 
The work of CL and JS is supported in part by the Gordon and Betty Moore Foundation, GBMF12234 and grant DOI 10.37807/gbmf12234. 
JS is also supported by the Department of Energy, Office of Science, under Award No. DE-SC0007859. 
PS is funded by the Istituto Nazionale di Fisica Nucleare (INFN) through the project of the InDark INFN Special Initiative: ``Neutrinos and other light relics in view of future cosmological observations'' (n.23590/2021). 
PS would like to thank the organizers of CATCH22+2 for their hospitality. 

%--------------------------------------------------------------------
\bibliography{main}

\end{document}